\documentclass[aps,prl,reprint,superscriptaddress]{revtex4-2}
\usepackage[utf8]{inputenc}
\usepackage[T1]{fontenc}
\usepackage{amsmath, amssymb, graphicx, lmodern, braket}

\usepackage{microtype} 

\usepackage{url}

\usepackage[colorlinks=true,linkcolor=blue,citecolor=blue]{hyperref}

 \usepackage[dvipsnames]{xcolor}

\newcommand*{\tn}[1]{\textnormal{#1}}

\newcommand*{\unit}[1]{\,\mathrm{#1}}

\newcommand{\vsd}{V_\tn{sd}}
\newcommand{\vtg}{V_\tn{tg}}

\newcommand{\vx}{V_\tn{ex}}
\newcommand{\vcn}{V_\tn{cn}}
\newcommand{\vcp}{V_\tn{cp}}

\newcommand{\isd}{I_\tn{sd}}
\newcommand{\gsd}{G_\tn{sd}}
\newcommand{\gmean}{G_\tn{mean}}

\newcommand{\tneel}{T_\tn{N}}

\newcommand{\bsat}{H_\mathrm{sat}}

\newcommand{\er}{\varepsilon_\mathrm{r}}

\newcommand{\suppmat}{SM}\newcommand{\polymer}{polymer layer}
\newcommand{\papertitle}{Efficient non-volatile electric control of magnetosensing in CrSBr}

\newcommand{\up}{\textit{up}}
\newcommand{\down}{\textit{down}}
 
\begin{document}

\title{\papertitle}
\newcommand{\fiztsz}{Department of Physics, Institute of Physics, Budapest University of Technology and Economics, Muegyetem rkp. 3., 1111 Budapest, Hungary}

\newcommand{\lenduletaffil}{MTA-BME Correlated van der Waals Structures Momentum Research Group, M\H{u}egyetem rkp. 3., H-1111 Budapest, Hungary}

\newcommand{\scaffil}{MTA-BME Superconducting Nanoelectronics Momentum Research Group, M\H{u}egyetem rkp. 3., H-1111 Budapest, Hungary}

\newcommand{\skaffil}{Institute of Quantum Science, Department of Physics, Inha University, 22212 Incheon, South Korea}

\newcommand{\cicaffil}{CIC nanoGUNE BRTA, 20018 Donostia--San Sebastian, Basque Country, Spain}

\newcommand{\ehuaffil}{Departamento de Polímeros y Materiales Avanzados: Física, Química y Tecnología, Facultad de Químicas (EHU), Apartado 1072, 20080, San Sebastián, Spain}

\newcommand{\cfmaffil}{Centro de Física de Materiales--Materials Physics Center (CFM-MPC), CSIC--UPV/EHU, Donostia--San Sebastián, 20018, Spain}

\newcommand{\ikeraffil}{Ikerbasque, Basque Foundation for Science, 48009 Bilbao, Spain}

\newcommand{\rugaffil}{Zernike Institute for Advanced Materials, University of Groningen, Groningen, Netherlands}

\newcommand{\isisaffil}{University of Strasbourg \& CNRS, ISIS \& icFRC, 8 allee Gaspard Monge, 67000 Strasbourg, France}

\newcommand{\inlaffil}{INL - International Iberian Nanotechnology Laboratory, 4715-330 Braga, Portugal}

\author{B\'alint \textsc{F\"ul\"op}}
\email{fulop.balint@ttk.bme.hu}
\affiliation{\fiztsz}

\author{Yeonsu \textsc{Jeong}}
\affiliation{\skaffil}

\author{Sofia \textsc{Ferreira-Teixeira}}
\affiliation{\cicaffil}
\affiliation{\inlaffil}

\author{Covadonga \textsc{\'Alvarez-Garc\'ia}}
\affiliation{\cicaffil}
\affiliation{\ehuaffil}

\author{Tam\'as \textsc{Prok}}
\affiliation{\fiztsz}
\affiliation{\lenduletaffil}
\affiliation{\rugaffil}

\author{Xiaomin \textsc{Guo}}
\affiliation{\cicaffil}

\author{Endre \textsc{T\'ov\'ari}}
\affiliation{\fiztsz}
\affiliation{\lenduletaffil}

\author{Jong Hoon \textsc{Jung}}
\affiliation{\skaffil}

\author{Luis \textsc{Hueso}}
\affiliation{\cicaffil}
\affiliation{\ikeraffil}

\author{Paolo \textsc{Samori}}
\affiliation{\isisaffil}

\author{Marco \textsc{Gobbi}}
\email{marco.gobbi@ehu.eus}
\affiliation{\cfmaffil}
\affiliation{\ikeraffil}

\author{P\'eter \textsc{Makk}}
\email{makk.peter@ttk.bme.hu}
\affiliation{\fiztsz}
\affiliation{\lenduletaffil}

\author{Szabolcs \textsc{Csonka}}
\affiliation{\fiztsz}
\affiliation{\scaffil} \date{\today}

\begin{abstract}
In two-dimensional crystals, electrostatic gating can modulate the charge carrier density, thereby tuning their electrical properties.
In magnetic materials such as CrSBr, this approach can also enable direct electrical control of magnetoresistance.
Non-volatile gating is particularly attractive since it allows the carrier density to be maintained at a desired level without the need for a continuous gate bias, offering significant advantages for technological applications.
Here, we demonstrate a simple device architecture in which a solution-processed ferroelectric P(VDF-TrFE) layer is integrated directly onto a CrSBr channel, enabling non-volatile control of its magnetoresistance.
We investigate the magnetotransport response over a broad temperature range and demonstrate a reproducible and robust gate dependence.
Notably, the ferroelectric gate achieves a gating efficiency approximately one order of magnitude higher than that of a conventional SiO$_2$ dielectric, highlighting the potential of ferroelectric gating for electrical control of 2D magnetic materials.
\end{abstract} \maketitle

\section{Introduction}
\begin{figure*}
\centering
\includegraphics[width=\textwidth]{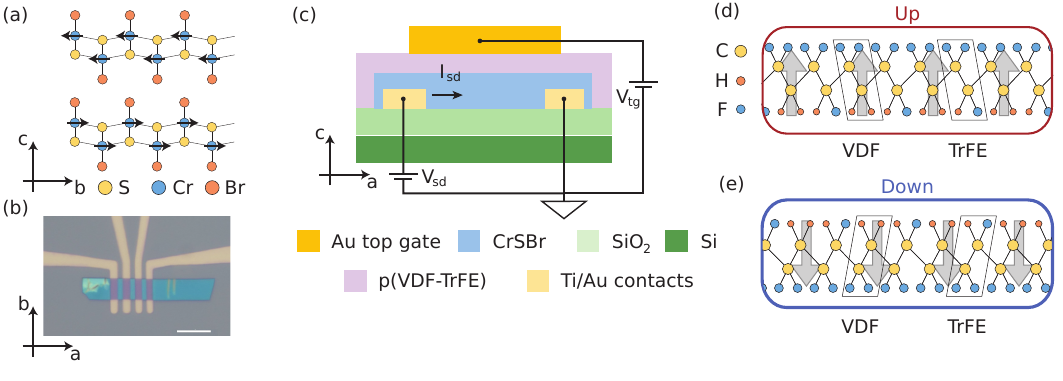}
\caption{
\textbf{Ferroelectric gating of CrSBr.}
(a)
Schematic representation of the crystal structure of CrSBr in the c--b plane with the arrangement of magnetic moments in the AFM phase.
(b)
Optical micrograph of the eight-layer-thick CrSBr flake (green) with the metallic contacts (golden leads) before polymer coating, lying on the Si/SiO$_2$ substrate (gray).
The crystallographic axes are marked in the corner, the flake size is $6 \unit{\mu m} \times 40 \unit{\mu m}$, the scale bar is 10\,\textmu{}m.
(c)
Schematic side view of the device.
The CrSBr flake is supported by a Si/SiO$_2$ substrate and contacted via regular evaporated Ti/Au contacts.
The top gate dielectric is a spin-coated layer of p(VDF-TrFE), which hosts the dipole moments that regulate the charge carrier density in the CrSBr flake.
(d-e)
Orientation of the molecular structure in the \polymer{} in the \textit{up} and \textit{down} states.
The orientation of the electric depolarization field is indicated by gray arrows.
}
\label{fig:schema}
\end{figure*}

The discovery of two-dimensional (2D) magnetic materials and the advent of \textit{van der Waals} heterostructures have opened new opportunities and potential applications in various fields such as spintronics, magnonics, spin-caloritronics, and materials science \cite{Gibertini2019, Ningrum2020, Jia2025}.
Specifically, antiferromagnets (AFMs) as building blocks possess several benefits over ferromagnets (FMs), e.g., their robustness against magnetic perturbation, fast spin dynamics, and lack of stray fields \cite{Baltz2018, Sierra2021, DalDin2024}.
In electric transport, antiferromagnets exhibit large magneto-resistance (MR) \cite{Song2018a, Klein2018}. This can strongly depend on the charge carrier density, which can be controlled in exfoliated thin flakes by using a gate electrode\cite{Wu2023}.
This electric control over the magnetic behaviour is an important feature for potential applications, because unlike operating with magnetic fields, electric gating can be reduced in size and power consumption and allow a much faster operation.
Intensive research is concentrated on this property in various systems \cite{Ohno2000, Weisheit2007, Maruyama2009, Chiba2011, Heron2014}, however, the need of a constant supply of a stable gate voltage can be a limiting factor in future applications.
Non-volatile gate control of the magnetic material can overcome this difficulty, but it typically involves a complex, multi-component design \cite{Yang2022}.
However, simple architectures can be achieved by the molecular coating of 2D interfaces\cite{Jo2022}, which also and opens the way towards multifunctional and highly tunable spintronic devices\cite{Meng2025}.
In this paper, we demonstrate these benefits by using a spin-coated ferroelectric polymer layer as gate dielectric in combination with a layered van der Waals antiferromagnet.

The 2D semiconductor CrSBr has emerged in the last few years and gained scientific attention due to its air stability, relatively high magnetic ordering temperatures and a strong coupling between electric and magnetic properties \cite{Telford2020, Lee2021, Wu2022, Ye2022, Jo2024, Lin2024, Jo2025, Liu2025, LopezPaz2022}.
The magnetic phases of CrSBr are attributed to the alignment of the magnetic moments of the Cr$^{3+}$ atoms in the $S=3/2$ spin state \cite{Goeser1990}.
Below the paramagnetic phase at high temperature, intralayer ferromagnetic correlations appear with a Curie temperature of ca.\ 160\,K in bulk and decreasing with decreasing number of layers down to 145\,K for the monolayer \cite{Telford2020, Lee2021}.
Magnetic moments align in the plane, along the crystallographic $b$ axis, while the other in-plane direction ($a$) is an intermediate magnetic axis, and the out-of-plane ($c$) direction is a hard axis.
Interlayer antiferromagnetic (AFM) coupling emerges between the ferromagnetic layers below the Néel temperature $\tneel = 132 \unit{K}$ in bulk \cite{Goeser1990}, which is only slightly affected by the layer number in thin samples \cite{Telford2020}.
In this state, individual layers retain their ferromagnetic alignment but the magnetic moment points in opposite direction in neighbouring layers corresponding to an A-type antiferromagnetic structure, depicted in Fig.\,\ref{fig:schema}a.
The highly anisotropic electric transport in CrSBr was intensively studied \cite{Telford2020, Wu2022}, moreover, the feature-rich MR was found to exhibit a strong gate-dependence \cite{Telford2022}.

Among polymer ferroelectrics, poly(vinylidene fluoride-co-trifluoroethylene), in short p(VDF-TrFE), is particularly attractive because it combines a low energy barrier for stabilizing the polar $\beta$ phase with excellent solution processability \cite{Naber2005, Naber2010}.
The incorporation of trifluoroethylene (TrFE) units into poly(vinylidene fluoride) (pVDF) suppresses the formation of non-polar crystalline phases, enabling robust room-temperature ferroelectricity through spontaneous $\beta$-phase crystallization.
Its ferroelectricity arises from the collective alignment of molecular dipoles associated with the highly polar C--F bonds along the polymer backbone.
The resulting remanent polarization produces a non-volatile internal electric field capable of efficiently modulating the carrier density of adjacent 2D materials without requiring continuous applied voltage, thus, power consumption.
Accordingly, p(VDF-TrFE) has been extensively adopted as a ferroelectric gate dielectric for non-volatile electronic and optoelectronic devices based on 2D semiconductors \cite{Jeong2026, Qiu2026}.
However, its application to non-volatile electrical control of 2D magnetic materials has remained largely unexplored.

In this paper, the ferroelectric control of the MR by charge carrier density in CrSBr is presented.
We show a ferroelectric behaviour through a hysteretic gate response.
Moreover, we demonstrate a large tunability of the MR using the ferroelectric polymer without requiring the constant application of gate voltage.
The reproducibility of the ferroelectric gating effect is shown and we estimate that top gating across the ferroelectric dielectric is an order of magnitude more powerful than back gating across the SiO$_2$ layer of the substrate.
Our results show that the combination of 2D magnetic materials and ferroelectric dielectrics can lead to an efficient, non-volatile electric control of magnetic signals in MR sensors.
 
\section{Results and Discussion}
Our sample design is based on a mechanically exfoliated eight-layer-thick CrSBr flake placed on pre-patterned Ti/Au contacts on a SiO$_2$ substrate, as shown in Fig.\,\ref{fig:schema}b.
It is known that the exfoliated flakes display an elongated shape along the magnetic intermediate ($a$) axis.
This cleaving anisotropy makes it easy to identify the crystallographic axes using optical microscopy, as indicated by the axes in the image.

A 0.5\,\textmu{}m thick p(VDF-TrFE) layer was used as a top gate dielectric on CrSBr and was covered by a top gate electrode made of Au, as shown in Fig.\,\ref{fig:schema}c.
Using the voltage of the top gate electrode ($\vtg$), the electric dipoles that form between the hydrogen and fluorine atoms in the polymer layer can be oriented, as illustrated in Fig.~\ref{fig:schema}d-e.
The dipoles retain their state even after grounding the top gate electrode, leading to a non-volatile gating effect.
The strong electric field stemming from the dipoles leads to the doping of the underlying CrSBr layer.
In addition to the change in the sample conductance, the large MR exhibited by CrSBr is also modified since it depends strongly on the charge carrier density.

\begin{figure}
\includegraphics[width=0.5\textwidth]{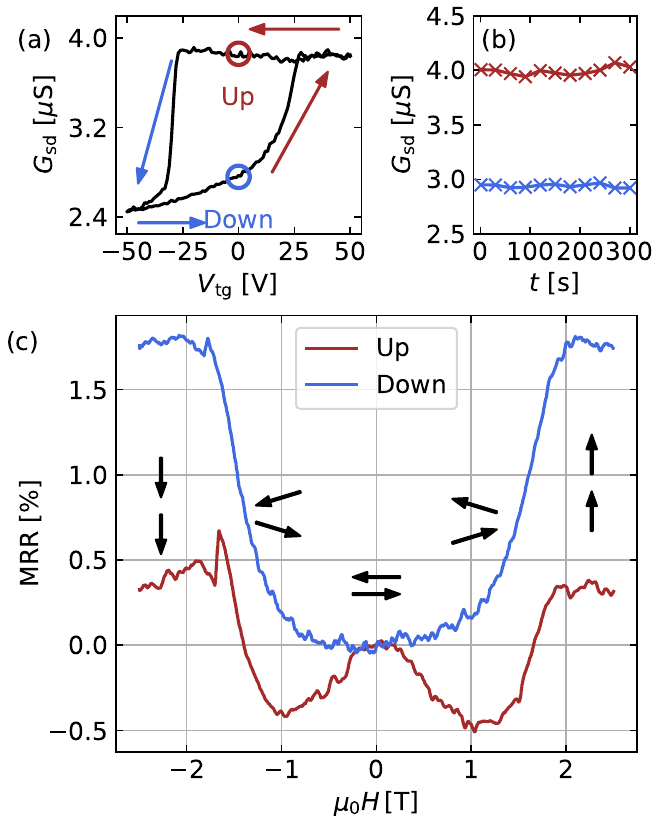}
\caption{
\textbf{Dipole state dependent magnetoresistance of CrSBr.}
(a)
Example of a transconductance measurement on a test device for $\vsd = 1 \unit{V}$ at room temperature.
$\gsd = \isd / \vsd$ is the sample conductance.
To set the \up{} state, $\vtg$ is increased above the coercive voltage of the layer, then decreased back to zero.
The ferroelectric polymer retains its polarization which keeps the conductance at the high-voltage level.
Setting the \down{} state requires $\vtg$ to decrease below the negative coercive voltage followed by setting it to zero.
(b)
Retention test of the dipole state on another test device, i.e., $\gsd$ vs.\ elapsed time measured at $\vtg = 0\unit{V}$ after setting the \up{} (\down) state at $\vtg = 50\unit{V}$ ($-50\unit{V}$).
(c)
Example curves of magneto-resistance ratio (MRR) vs.\ magnetic field along the $c$ axis corresponding to the two dipole states at $15\unit{K}$ measured on the main device.
Pairs of black arrows illustrate the sublattice magnetizations, where the horizontal direction corresponds to the $b$ axis, while the vertical to $c$.
The raw zero-field resistance is $235\unit{k\Omega}$ and $250\unit{k\Omega}$ for the \up{} and \down{} curves, respectively.
}
\label{fig:mrr}
\end{figure}

The preparation of the ferroelectric polymer dipole state is done at room temperature inside a variable temperature insert (VTI) by applying a gate voltage.
We found that the polymer can sustain dozens of such gating cycles without any sign of fatigue.
One example is presented in Fig.~\ref{fig:mrr}a in a transconductance measurement setup, where the source--drain conductance $\gsd = \isd/\vsd$ is plotted against the applied top gate voltage $\vtg$.
The source--drain voltage $\vsd$ was set to 1\,V throughout this paper, while the current $\isd$ was measured.
Increasing the gate voltage in the positive direction above the positive coercive voltage $\vcp$ flips the ligands around the polymer chain, thus, positively charged H atoms move towards the CrSBr flake.
Since the electric dipole vectors point upwards, as illustrated by gray arrows in Fig.\,\ref{fig:schema}d, we name this the \textit{\up} state of the \polymer.
This arrangement of dipoles induces an additional n-doping in the CrSBr layer, increasing the charge carrier density in the conduction band, and the sample conductance $\gsd$ \cite{Wu2022}, as shown in Fig.\,\ref{fig:mrr}a.
Then, decreasing the gate voltage to zero does not change $\gsd$ substantially.
In contrast, sweeping the top gate voltage in the negative direction beyond the negative coercive voltage $\vcn$, as shown in Fig.\,\ref{fig:schema}e, the top gate electrode flips the dipoles in the opposite (\textit{\down}) state, leading to a drop in the charge carrier density and in $\gsd$, as well.
As shown in Fig.~\ref{fig:mrr}b, the conductance of the CrSBr does not change over time after the gating cycle for either of the dipole states.

After preparing the \up{} or \down{} dipole state at room temperature as described above, the sample was cooled down in zero magnetic field at $\vtg = 0$ to select temperatures.
The polarization state of the \polymer{} feezes during cooling, therefore, the doping level of the CrSBr is fixed at the temperatures of the study.
After stabilizing the system at a target temperature, MR measurements were performed by recording $\isd$ as a function of magnetic field $H$ with $\vsd$ of 1\,V.
In the following, we focus on measurements in magnetic field in the out-of-plane direction, which corresponds to the magnetic hard axis $c$.
We analyse the magneto-resistance ratio (MRR) defined as $\tn{MRR}(H) = \left(R(H)-R_0\right)/R_0)$, where $R(H) = \vsd / \isd(H)$ is the two-terminal sample resistance, and $R_0 = R(H\!=\!0)$.
Magnetic field dependence along other directions and measurements on a second, qualitatively similar but four-layer-thick sample are shown in the Supporting Material (\suppmat).

\begin{figure*}
\includegraphics[width=\textwidth]{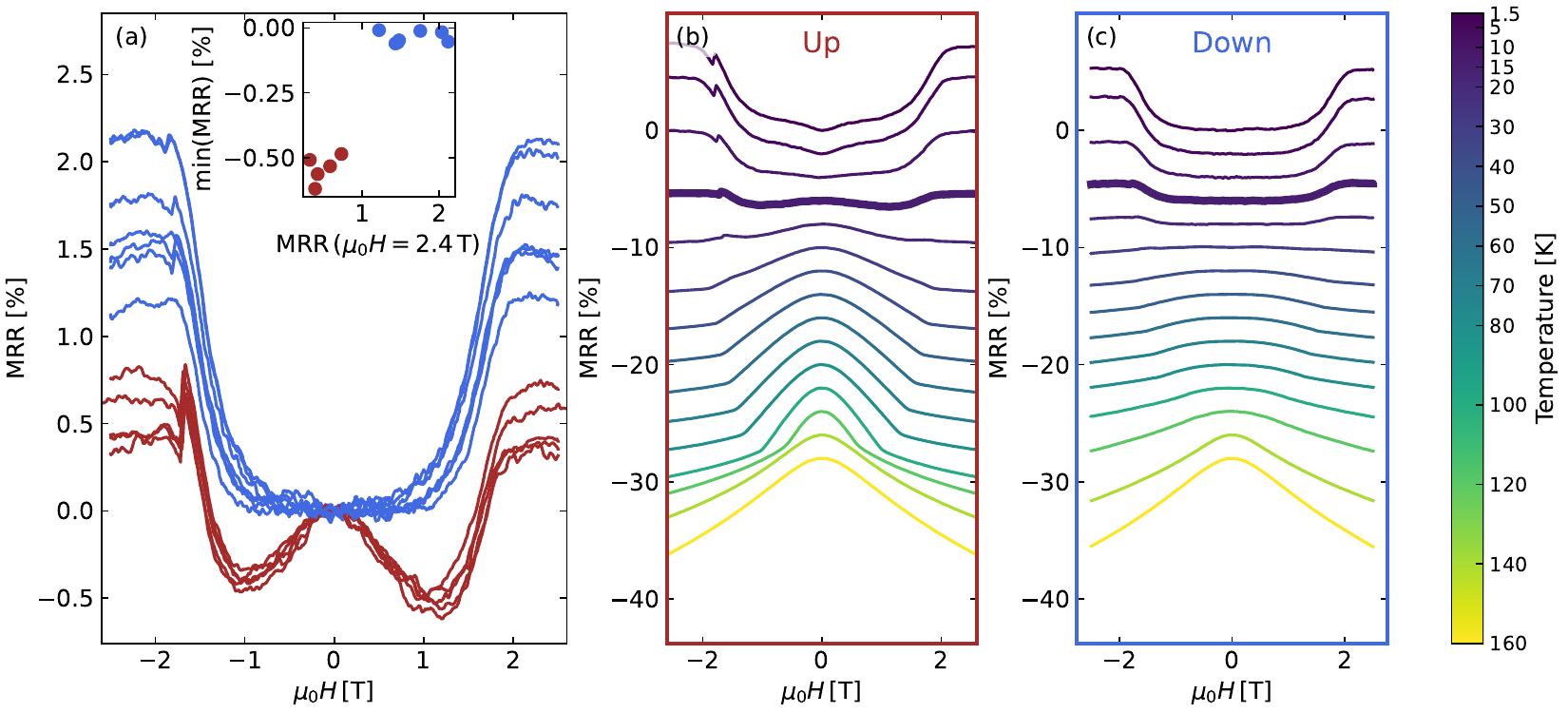}
\caption{
\textbf{Robustness of the MRR features.}
(a)
Two sets of MRR curves showing the markedly different magneto-resistance for the \up{} (red) and \down{} (blue) dipole states at $15\unit{K}$ along axis $c$.
Inset:
Scatter plot displaying the minimum value and the high field value of the MRR for the curves shown in the main panel.
(b-c)
MRR curves for a wide range of temperature values for the \up{} (b) and \down{} (c) dipole states shifted for clarity.
Each MRR curve was recorded at a fixed temperature indicated on the rightmost colorbar, where the labels correspond to the temperatures of the curves.
}
\label{fig:reprod}
\end{figure*}

In Fig.\,\ref{fig:mrr}c, the MRR curves of the two dipole states are compared.
After setting the up state at room temperature, the sample was cooled down to 15\,K and the red curve was measured.
After warming up again and setting the down state, the sample was again cooled down to 15\,K and we obtained the blue curve.
Black arrows represent the spin orientation of neighbouring layers for different magnetic fields.
In zero magnetic field, the AFM coupling between neighbouring layers leads to opposite spins along the $b$ axis (easy axis).
Increasing an external magnetic field along the $c$ axis leads to both spins canting towards this direction.
In this canted regime, carrier density dependent scattering processes lead to different MRR values for the two dipole states.
In the \up{} state, the measurement exhibits a considerable negative MRR (nMRR).
Increasing the magnetic field over a transition value of ca.\ 1\,T, the MRR starts to increase, leading to a positive MRR (pMRR) contribution to the signal.
Whereas in the \down{} state, the nMRR component is absent, and the MRR shows a monotonous increasing trend up to a saturation field of $\mu_0 | H_\tn{sat} | \approx 2 \unit{T}$.
At higher magnetic fields, both curves approximately saturate, indicating that spins reach 90$^\circ$ rotation and the layer magnetization will fully align with the external field, entering a spin-polarized FM state.
Here as well, the MRR signal of the two dipole states differ significantly and is lower in the \up{} state.
Gate-tunable MRR signal on a monolayer CrSBr was previously reported, however, with an opposite trend in function of applied gate voltage \cite{Telford2022}.
This may be attributed to a difference in flake thickness (monolayer vs.\ 8 layers) as well as in the device buildup (top gate vs.\ back gate).

In order to investigate the reproducibility of this ferroelectric gating effect, we performed several cycles of dipole flips and subsequent MR measurements.
Fig.\,\ref{fig:reprod}a shows a series of MRR curves at 15\,K with the magnetic field ramping in the negative direction.
We extracted the minimal value of MRR and the MRR value at $2.4\unit{T}> \mu_0 \bsat$.
This pair of values for each MRR curve is plotted in the inset of Fig. 3a as a single dot, colored corresponding to the dipole state.
This representation demonstrates the difference between the two states and their reproducibility at 15\,K.

A qualitative difference between the MRR curve shapes is present at all investigated temperatures up to $\tneel$, as shown in Fig.\,\ref{fig:reprod}b-c for the \up{} and \down{} states, respectively.
The curves corresponding to 15\,K presented in panel (a) are plotted with thick lines for comparison.
Below this temperature, the \up{} state features an overall pMRR behaviour.
At 15\,K, the nMRR component appears in this state, and above 20\,K, its contribution is dominant and the curve does not show a distinct pMRR region.
Regarding the MRR curves of the \down{} state, we can observe the dominant line shape to change from pMRR to nMRR here as well, although this occurs at a higher temperature (around 30\,K), and the overall size of the nMRR signal stays smaller than in the \up{} case.
At magnetic fields higher than $\bsat$ corresponding to the ferromagnetic state, both dipole states exhibit increasing nMRR with increasing temperature.
This feature persists even above $\tneel$, i.e., in the paramagnetic phase.
These measurements clearly present that the MRR curve of the CrSBr channel is characteristic of the \up{} or \down{} state of the ferroelectric \polymer.

\begin{figure*}
\includegraphics[width=\textwidth]{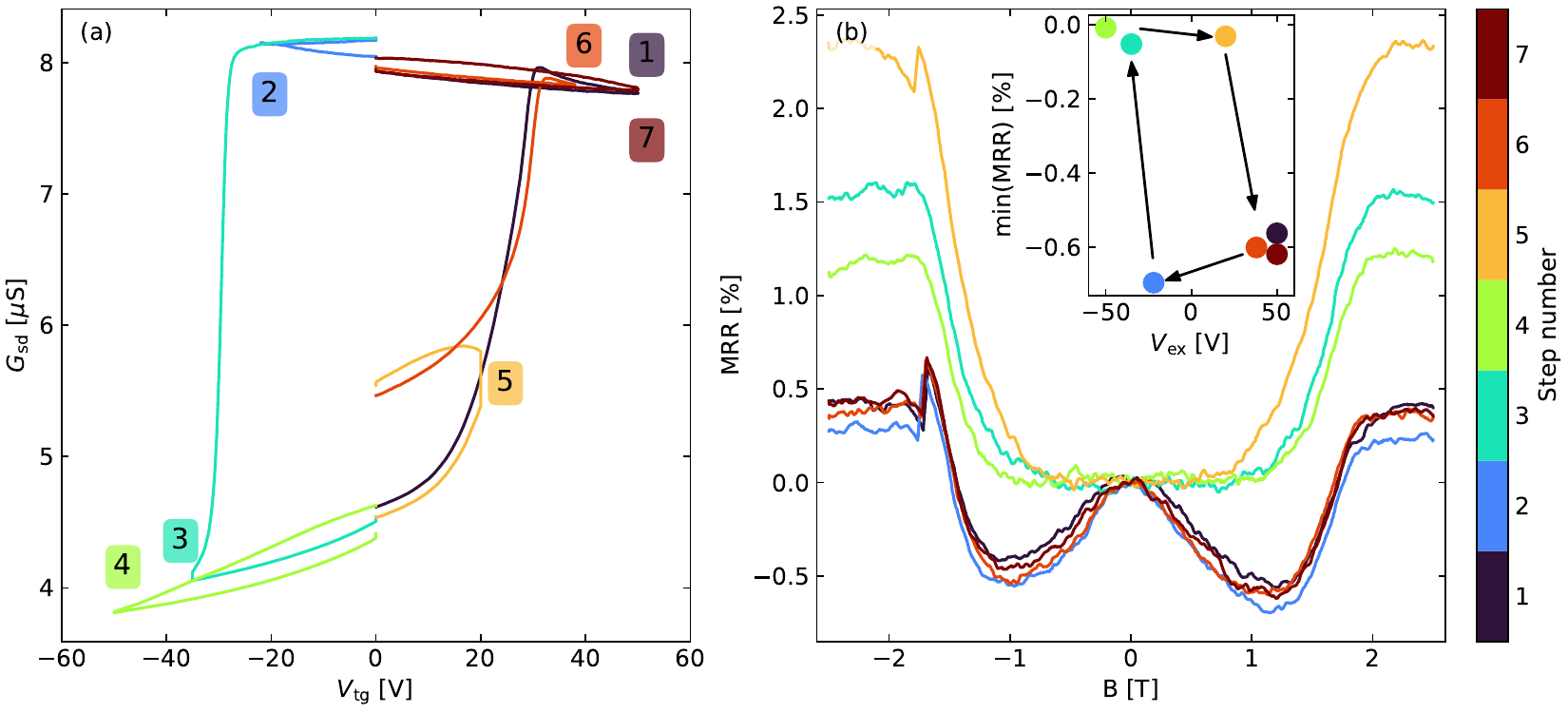}
\caption{
\textbf{Coercivity of the ferroelectric layer.}
(a)
Summary of seven transconductance measurements at room temperature with various extremal voltages $\vx$.
In each case, $\vtg$ is ramped from zero to $\vx$, then decreased back to zero.
(b)
Corresponding MRR curves measured at 15\,K after each measurement showed in panel (a). 
The curves group around the two shapes corresponding to the \up{} and \down{} states shown in Fig.\ref{fig:mrr}.
Inset: 
The minimum value of each MRR curve in panel (b) vs.\,$\vx$.
}
\label{fig:partial}
\end{figure*}

To test the coercivity and robustness of the \polymer, we performed a series of gating cycles with different extremal top gate voltage $\vx$, as presented in Fig\,\ref{fig:partial}.
Each cycle consisted of a transconductance measurement at room temperature, sweeping $\vtg$ to $\vx$ and back to 0, shown in panel (a), followed by a series of MR measurements at low temperatures, from which the 15\,K data is shown in  panel (b).
The corresponding curves on both panels and the steps described below are color coded according to the side bar.
In step 1, the \polymer{} was initially in the low conductance \down{} state.
Ramping $\vtg$ to $\vx = + 50\unit{V}$ brings it into the fully polarized \up{} state.
Meanwhile, the conductance increases rapidly, then stays approximately constant as $\vtg$ sweeps back to zero, in a similar way to Fig.\,\ref{fig:mrr}a.
The corresponding MRR curve is also similar to the one presented as the \up{} state in Fig.\,\ref{fig:mrr}c.
In step 2, a cycle with $\vx = -22\unit{V}$ is performed, therefore $\vx$ stays above the $\vcn$ negative coercive voltage observed for this sample.
The conductance stays close to the value of the \up{} state throughout this step, furthermore, the MRR curve also overlaps the previous one, indicating the \polymer{} staying in the \up{} state.
As a next step, as we applied $\vx = -35\unit{V}$, a value slightly beyond $\vcn$, we observed a drop in sample conductance and the corresponding MRR curve also exhibits features close to that of the \down{} state with a lack of the nMRR component and elevated high field MRR value.
These features indicate a change in the dipole state in the \polymer.
To reach the fully polarized \down{} state, we applied $\vx = -50 \unit{V}$ in step 4.
The sample conductance decreased and the high-field MRR increased even further with an absent nMRR.
In step 5, $\vx = +20\unit{V}$ was applied, which is little under the $\vcp$ positive coercive voltage.
This brought the system to an intermediate state, where the MRR curve still clearly exhibits the features of the \down{} state, meanwhile, the conductance increases substantially with $\vtg$ and its final value is somewhat higher than in the previous step.
In step 6, both the transconductance curve and the MRR indicate that the employed $\vx = +38\unit{V} > \vcp$ caused the onset of the \up{} state, after which applying $\vx = +50\unit{V}$ in step 7 did not cause any further noticeable change in the polymer state.

\begin{figure}
\centering
\includegraphics[width=0.48\textwidth]{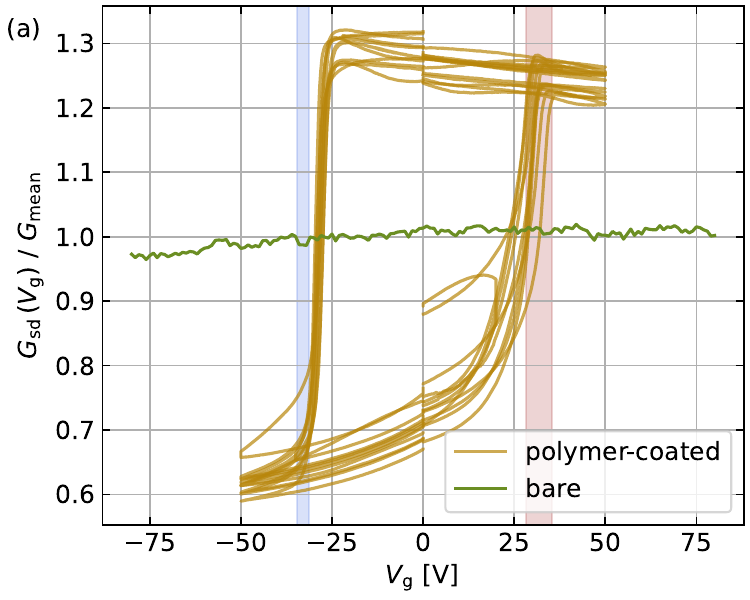}
\includegraphics[width=0.48\textwidth]{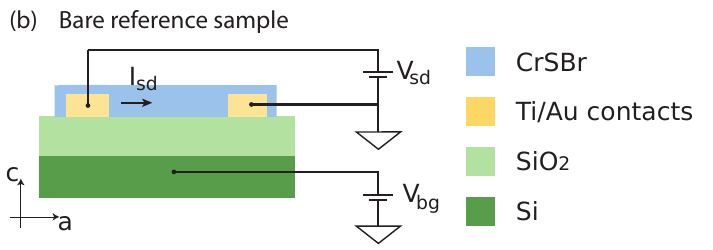}
\caption{
\textbf{Comparison of gate efficiency.}
(a)
Source--drain normalized conductance $\gsd$ of the presented device in response to the top gate (gold curves) compared to that of a reference on a bare sample (green curve), i.e., without the \polymer{}, in response to its back gate.
Both measurements were performed at $\vsd = 1\unit{V}$ and room temperature, both dataset normalized by their mean conductance.
The shaded blue and red areas indicate the coercive voltages $\vcn$ and $\vcp$, respectively.
(b)
Schematic side view of the bare reference sample.
Here, the conventional Si/SiO$_2$ back gate was used.
}
\label{fig:allgatesweeps}
\end{figure}

Based on these measurements, one can see that the MR of the underlying CrSBr flake clearly follows the ferroelectric state of the \polymer, which is summarized in the inset of Fig.\,\ref{fig:partial}b.
Here, the minimal MRR value of the 15\,K MR measurement is plotted against the $\vx$ used to prepare the polymer state.
The data shows that the polymer consistently retains its dipole state over subsequent thermal cycles until the top gate voltage exceeds a well defined coercive voltage of the opposite sign.
In order to determine the values of the coercive voltage in the two directions, we analyzed 25 transconductance measurements as plotted in Fig.\,\ref{fig:allgatesweeps}a.
Here, the conductance values are normalized by $\gmean=6.1\unit{\mu S}$, which is the mean of all the measurement points.
The overlap of the lines indicate a highly reproducible polymer behaviour.
Extracting the coercive voltage (see \suppmat for details) yielded $\vcn = -32.9 \pm 1.7 \unit{V}$ and  $\vcp = 31.8 \pm 3.5 \unit{V}$ for the negative and positive direction, respectively, as indicated in Fig.\,\ref{fig:allgatesweeps}a by the two shaded areas.
The estimated values of coercive voltage in the two directions overlap from curve to curve within the margin of error, however, on the negative $\vtg$ side, the shape of the downturn of the conductance curve is noticeably sharper than the upturn on the positive side.
Moreover, the slope of the conductance curve during $\vtg$ sweeping from $+50\unit{V}$ to zero is markedly different from the slope when sweeping from $-50\unit{V}$ to zero.
These observations suggest that additional charge traps can appear at the polymer interface depending on the dipole state which can partially screen the effect of the top gate electrode\cite{Park2022}.
The thickness of the polymer layer was $d = 0.5\,\unit{\mu m}$, therefore, the corresponding coercive electric field of the \polymer{} is roughly $65\unit{mV/nm}$, which is slightly smaller than in previous works \cite{Neese2008, Liang2023}.

To demonstrate the high gating efficiency, we also performed a transconductance measurement at room temperature on a bare reference sample, built on a CrSBr flake with similar thickness as the coated one, but without any capping layer, depicted in the \suppmat.
In case of the bare sample, the gating was achieved by using the conventional doped Si back gate across the 300\,nm thick SiO$_2$ layer of the substrate, as shown in Fig.\,\ref{fig:allgatesweeps}b.
In Fig.\,\ref{fig:allgatesweeps}a, a comparison is shown of top-gated transconductance curves measured on the polymer-coated (gold) and the back-gated bare samples (green).
Both the top and the back gate conduction datasets are normalized by their respective mean conduction value to account for the different aspect ratio of the flakes.
Flipping the dipole state of the polymer-coated sample from \down{} to \up{} induces a roughly $100\%$ increase in the conductance.
In contrast, by ramping the back gate voltage of the bare sample from $-80\unit{V}$ to $+80\unit{V}$, $\gsd$ merely changes by $5\%$.
We note that the change in the conductance is not directly proportional to the change in density due to the density dependent scattering mechanisms and, possibly, screening effects, but we assume the dependence is monotonic\cite{Wu2022}.
Therefore, this comparison clearly indicates a large difference between the polymer and the SiO$_2$ dielectrics.

We estimated the relative permittivity of the polymer from independent capacitance measurements on a layer of the same thickness (see \suppmat), and found $\er = 7.6 \pm 0.4$.
The polarization density measured in the \up{} and \down{} states is around $P_0 = \pm$4\,\textmu{}C\,cm$^{-2}$, which corresponds to $\pm E_0 = P_0\,d/\er \varepsilon_0 = \pm 600\unit{mV\,nm^{-1}}$ depolarization field, where the two signs correspond to the two dipole states (see details in the \suppmat).
Changing the polarization state from \up{} to \down{} implies a charge carrier density change in the CrSBr of $2 P_0 / e = 5\cdot 10^{13}\unit{cm^{-2}}$, where $e$ is the elementary charge.
As a comparison, $700\unit{V}$ would be needed on the back gate electrode with a 300\,nm SiO$_2$ dielectric (using a parallel-plate capacitor model with $\er$ = 3.9) to obtain the same amount of charge density change.
Dividing by $\vcp - \vcn \approx 60\unit{V}$ means a roughly $10\times$ higher efficiency in the case of the polymer.
This is unattainable with traditional dielectrics such as SiO$_2$ or hexagonal boron nitride due to dielectric defects and breakdown.
 
\section{Conclusion}
In conclusion, we demonstrated the ferroelectric gating of CrSBr using a p(VDF-TrFE) dielectric and investigated the gating effect on the MR signal of the conductive channel.
The presence of the p(VDF--TrFE) layer not only enables a non-volatile gating in the system, but it also allows to tune the charge carrier density in CrSBr to an extent far beyond what is reachable with conventional dielectrics, due to its one order of magnitude higher gating efficiency.
Our device geometry and proof of concept paves the way towards a simplified manufacturing of non-volatile electric gating and magnetic sensing applications.
 
\section{Acknowledgments}
This work was supported by 
the FLAG-ERA grant MULTISPIN, 
by MICIU/AEI/10.13039/501100011033 (Grant CEX2020-001038-M), 
by MICIU/AEI and ERDF/EU (PCI2021-122038-2A and Projects PID2024-157558NB-C22 and PID2024-155708OB-I00), 
and by INHA UNIVERSITY Research Grant. 
M.G.\ acknowledges support from MICIU/AEI and the European Union NextGenerationEU/PRTR (RYC2021-031705-I). 
S.F.-T.\ acknowledges financial support from the European Union via Marie Skłodowska-Curie grant agreement number 101106104 CHEERS. 
E.T.\ acknowledges financial support from NKKP STARTING grant No.\ 150232.
P.S.\ acknowledges financial support from the European Union through the 2D-PRINTABLE project (GA-101135196), the Agence Nationale de la Recherche through the Interdisciplinary Thematic Institute SysChem via the IdEx Unistra (ANR-10-IDEX-0002) within the program Investissement d’Avenir, the Foundation Jean-Marie Lehn, and the Institut Universitaire de France (IUF).
 
\section{Author contributions}
S. F-T. C. Á-G, X. G., and Y.J. fabricated the samples with the guidance of M. G., L.H. and P. S.
The transport experiments were performed by B. F. and P. T.\quad
B. F., E. T., P. M., Sz. Cs. analysed the measurement data.
Y. J., and P. S., developed the solution process.
Y. J., and J. H. J. characterized the polymer.
The design was developed by M. G., P. S., Sz. Cs., P. M., and L. H.
The manuscript was prepared by B. F., P. M., and E. T., with input from all authors.
  
\onecolumngrid

\bibliography{cited}

@article{Song2018a,
	author = {Tiancheng Song and Xinghan Cai and Matisse Wei-Yuan Tu and Xiaoou Zhang and Bevin Huang and Nathan P. Wilson and Kyle L. Seyler and Lin Zhu and Takashi Taniguchi and Kenji Watanabe and Michael A. McGuire and David H. Cobden and Di Xiao and Wang Yao and Xiaodong Xu},
	journal = {Science},
	title = {Giant tunneling magnetoresistance in spin-filter van der Waals heterostructures},
	year = {2018},
	number = {6394},
	pages = {1214-1218},
	volume = {360},
	doi = {10.1126/science.aar4851},
	eprint = {https://www.science.org/doi/pdf/10.1126/science.aar4851},
	url = {https://www.science.org/doi/abs/10.1126/science.aar4851}
}

@article{Yang2022,
	author = {Yang, Hyunsoo and Valenzuela, Sergio O. and Chshiev, Mairbek and Couet, Sébastien and Dieny, Bernard and Dlubak, Bruno and Fert, Albert and Garello, Kevin and Jamet, Matthieu and Jeong, Dae-Eun and Lee, Kangho and Lee, Taeyoung and Martin, Marie-Blandine and Kar, Gouri Sankar and Sénéor, Pierre and Shin, Hyeon-Jin and Roche, Stephan},
	journal = {Nature},
	title = {Two-dimensional materials prospects for non-volatile spintronic memories},
	year = {2022},
	issn = {1476-4687},
	month = June,
	number = {7915},
	pages = {663--673},
	volume = {606},
	doi = {10.1038/s41586-022-04768-0},
	publisher = {Springer Science and Business Media LLC}
}

@article{Telford2020,
	author = {Telford, Evan J. and Dismukes, Avalon H. and Lee, Kihong and Cheng, Minghao and Wieteska, Andrew and Bartholomew, Amymarie K. and Chen, Yu-Sheng and Xu, Xiaodong and Pasupathy, Abhay N. and Zhu, Xiaoyang and Dean, Cory R. and Roy, Xavier},
	journal = {Adv. Mater.},
	title = {Layered Antiferromagnetism Induces Large Negative Magnetoresistance in the van der Waals Semiconductor CrSBr},
	year = {2020},
	issn = {0935-9648},
	month = sep,
	number = {37},
	pages = {2003240},
	volume = {32},
	publisher = {John Wiley \& Sons, Ltd},
	url = {https://doi.org/10.1002/adma.202003240}
}

@article{Heron2014,
	author = {Heron, J. T. and Bosse, J. L. and He, Q. and Gao, Y. and Trassin, M. and Ye, L. and Clarkson, J. D. and Wang, C. and Liu, Jian and Salahuddin, S. and Ralph, D. C. and Schlom, D. G. and Íñiguez, J. and Huey, B. D. and Ramesh, R.},
	journal = {Nature},
	title = {Deterministic switching of ferromagnetism at room temperature using an electric field},
	year = {2014},
	issn = {1476-4687},
	month = Dec,
	number = {7531},
	pages = {370--373},
	volume = {516},
	doi = {10.1038/nature14004},
	publisher = {Springer Science and Business Media LLC}
}

@article{Jo2024,
	author = {Jo, Junhyeon and Mañas-Valero, Samuel and Coronado, Eugenio and Casanova, Fèlix and Gobbi, Marco and Hueso, Luis E.},
	journal = {Nano Lett.},
	title = {Nonvolatile Electric Control of Antiferromagnet CrSBr},
	year = {2024},
	issn = {1530-6984},
	month = apr,
	number = {15},
	pages = {4471--4477},
	volume = {24},
	comment = {doi: 10.1021/acs.nanolett.4c00348},
	doi = {10.1021/acs.nanolett.4c00348},
	publisher = {American Chemical Society},
	url = {https://doi.org/10.1021/acs.nanolett.4c00348}
}

@article{Ye2022,
	author = {Ye, Chen and Wang, Cong and Wu, Qiong and Liu, Sheng and Zhou, Jiayuan and Wang, Guopeng and Söll, Aljoscha and Sofer, Zdenek and Yue, Ming and Liu, Xue and Tian, Mingliang and Xiong, Qihua and Ji, Wei and Renshaw Wang, Xiao},
	journal = {ACS Nano},
	title = {Layer-Dependent Interlayer Antiferromagnetic Spin Reorientation in Air-Stable Semiconductor CrSBr},
	year = {2022},
	issn = {1936-086X},
	month = may,
	number = {8},
	pages = {11876--11883},
	volume = {16},
	comment = {csb linear chain model},
	doi = {10.1021/acsnano.2c01151},
	publisher = {American Chemical Society (ACS)}
}

@article{Wu2023,
	author = {Wu, Fan and Gibertini, Marco and Watanabe, Kenji and Taniguchi, Takashi and Gutiérrez-Lezama, Ignacio and Ubrig, Nicolas and Morpurgo, Alberto F.},
	journal = {Advanced Materials},
	title = {Gate-Controlled Magnetotransport and Electrostatic Modulation of Magnetism in 2D Magnetic Semiconductor CrPS4},
	year = {2023},
	number = {30},
	pages = {2211653},
	volume = {35},
	doi = {https://doi.org/10.1002/adma.202211653},
	eprint = {https://advanced.onlinelibrary.wiley.com/doi/pdf/10.1002/adma.202211653},
	url = {https://advanced.onlinelibrary.wiley.com/doi/abs/10.1002/adma.202211653}
}

@article{Maruyama2009,
	author = {Maruyama, T. and Shiota, Y. and Nozaki, T. and Ohta, K. and Toda, N. and Mizuguchi, M. and Tulapurkar, A. A. and Shinjo, T. and Shiraishi, M. and Mizukami, S. and Ando, Y. and Suzuki, Y.},
	journal = {Nature Nanotechnology},
	title = {Large voltage-induced magnetic anisotropy change in a few atomic layers of iron},
	year = {2009},
	issn = {1748-3395},
	month = Jan,
	number = {3},
	pages = {158--161},
	volume = {4},
	doi = {10.1038/nnano.2008.406},
	publisher = {Springer Science and Business Media LLC}
}

@article{Goeser1990,
	author = {O. Göser and W. Paul and H.G. Kahle},
	journal = {Journal of Magnetism and Magnetic Materials},
	title = {Magnetic properties of CrSBr},
	year = {1990},
	issn = {0304-8853},
	number = {1},
	pages = {129-136},
	volume = {92},
	doi = {https://doi.org/10.1016/0304-8853(90)90689-N},
	url = {https://www.sciencedirect.com/science/article/pii/030488539090689N}
}

@article{Jo2022,
	author = {Jo, Junhyeon and Calavalle, Francesco and Martín-García, Beatriz and Tezze, Daniel and Casanova, Fèlix and Chuvilin, Andrey and Hueso, Luis E. and Gobbi, Marco},
	title = {Exchange Bias in Molecule/Fe3GeTe2 van der Waals Heterostructures via Spinterface Effects},
	journal = {Advanced Materials},
	year = {2022},
	volume = {34},
	number = {21},
	pages = {2200474},
	doi = {https://doi.org/10.1002/adma.202200474},
	eprint = {https://onlinelibrary.wiley.com/doi/pdf/10.1002/adma.202200474},
	url = {https://onlinelibrary.wiley.com/doi/abs/10.1002/adma.202200474}
}

@article{Liang2023,
	author = {Liang, Shanchuan and Xie, Ti and Blumenschein, Nicholas A. and Zhou, Tong and Ersevim, Thomas and Song, Zhihao and Liang, Jierui and Susner, Michael A. and Conner, Benjamin S. and Gong, Shi-Jing and Wang, Jian-Ping and Ouyang, Min and Žutić, Igor and Friedman, Adam L. and Zhang, Xiang and Gong, Cheng},
	journal = {Nature Electronics},
	title = {Small-voltage multiferroic control of two-dimensional magnetic insulators},
	year = {2023},
	issn = {2520-1131},
	month = mar,
	number = {3},
	pages = {199--205},
	volume = {6},
	comment = {fecsb, pvdf-trfe cgt},
	doi = {10.1038/s41928-023-00931-1},
	publisher = {Springer Science and Business Media LLC},
	refid = {Liang2023},
	url = {https://doi.org/10.1038/s41928-023-00931-1}
}

@article{Sierra2021,
	author = {Sierra, Juan F. and Fabian, Jaroslav and Kawakami, Roland K. and Roche, Stephan and Valenzuela, Sergio O.},
	journal = {Nature Nanotechnology},
	title = {Van der Waals heterostructures for spintronics and opto-spintronics},
	year = {2021},
	issn = {1748-3395},
	month = July,
	number = {8},
	pages = {856--868},
	volume = {16},
	doi = {10.1038/s41565-021-00936-x},
	publisher = {Springer Science and Business Media LLC}
}

@article{Lin2024,
	author = {Lin, Xiaohanwen and Wu, Fan and L\'opez-Paz, Sara A. and von Rohr, Fabian O. and Gibertini, Marco and Guti\'errez-Lezama, Ignacio and Morpurgo, Alberto F.},
	journal = {Phys. Rev. Res.},
	title = {Influence of magnetism on vertical hopping transport in CrSBr},
	year = {2024},
	month = {Feb},
	pages = {013185},
	volume = {6},
	comment = {csb linear regime},
	doi = {10.1103/PhysRevResearch.6.013185},
	issue = {1},
	numpages = {9},
	publisher = {American Physical Society},
	url = {https://link.aps.org/doi/10.1103/PhysRevResearch.6.013185}
}

@article{Wu2022,
	author = {Wu, Fan and Gutiérrez-Lezama, Ignacio and López-Paz, Sara A. and Gibertini, Marco and Watanabe, Kenji and Taniguchi, Takashi and von Rohr, Fabian O. and Ubrig, Nicolas and Morpurgo, Alberto F.},
	journal = {Advanced Materials},
	title = {Quasi-1D Electronic Transport in a 2D Magnetic Semiconductor},
	year = {2022},
	number = {16},
	pages = {2109759},
	volume = {34},
	comment = {csb},
	doi = {https://doi.org/10.1002/adma.202109759},
	eprint = {https://onlinelibrary.wiley.com/doi/pdf/10.1002/adma.202109759},
	url = {https://onlinelibrary.wiley.com/doi/abs/10.1002/adma.202109759}
}

@article{Meng2025,
	author = {Meng, Ke and Li, Min and Guo, Lidan and Zhang, Rui and Guo, Ankang and Liu, Mingzhe and Gu, Xianrong and Qin, Yang and Yang, Tingting and Yang, Xueli and Hu, Shunhua and Zhang, Cheng and Zheng, Ruiheng and Wu, Meng and Sun, Xiangnan},
	journal = {Advanced Materials},
	title = {Room-Temperature Organic Spintronic Devices with Wide Range Magnetocurrent Tuning and Multifunctionality via Electro-Optical Compensation Strategy},
	year = {2025},
	number = {11},
	pages = {2417995},
	volume = {37},
	doi = {https://doi.org/10.1002/adma.202417995},
	eprint = {https://advanced.onlinelibrary.wiley.com/doi/pdf/10.1002/adma.202417995},
	url = {https://advanced.onlinelibrary.wiley.com/doi/abs/10.1002/adma.202417995}
}

@article{Naber2005,
	author = {Naber, Ronald C. G. and Tanase, Cristina and Blom, Paul W. M. and Gelinck, Gerwin H. and Marsman, Albert W. and Touwslager, Fred J. and Setayesh, Sepas and de Leeuw, Dago M.},
	journal = {Nature Materials},
	title = {High-performance solution-processed polymer ferroelectric field-effect transistors},
	year = {2005},
	issn = {1476-4660},
	month = Feb,
	number = {3},
	pages = {243--248},
	volume = {4},
	doi = {10.1038/nmat1329},
	publisher = {Springer Science and Business Media LLC}
}

@proceedings{Qiu2026,
	title = {Ferroelectrics Hybrids: Harnessing Multifunctionality of 2D Semiconductors in the Post-Moore Era},
	year = {2026},
	number = {7},
	volume = {38},
	author = {Qiu, Haixin and Qian, Xiaoshi and Qian, Dahong and Samorì, Paolo},
	doi = {https://doi.org/10.1002/adma.202517269},
	eprint = {https://advanced.onlinelibrary.wiley.com/doi/pdf/10.1002/adma.202517269},
	journal = {Advanced Materials},
	pages = {e17269},
	url = {https://advanced.onlinelibrary.wiley.com/doi/abs/10.1002/adma.202517269}
}

@article{Chiba2011,
	author = {Chiba, D. and Fukami, S. and Shimamura, K. and Ishiwata, N. and Kobayashi, K. and Ono, T.},
	journal = {Nature Materials},
	title = {Electrical control of the ferromagnetic phase transition in cobalt at room temperature},
	year = {2011},
	issn = {1476-4660},
	month = Oct,
	number = {11},
	pages = {853--856},
	volume = {10},
	doi = {10.1038/nmat3130},
	publisher = {Springer Science and Business Media LLC}
}

@article{Jo2025,
	author = {Jo, Junhyeon and Suárez‐Rodríguez, Manuel and Mañas‐Valero, Samuel and Coronado, Eugenio and Souza, Ivo and de Juan, Fernando and Casanova, Fèlix and Gobbi, Marco and Hueso, Luis E.},
	journal = {Advanced Materials},
	title = {Anomalous Nonlinear Magnetoconductivity in van der Waals Magnet CrSBr},
	year = {2025},
	issn = {1521-4095},
	month = mar,
	number = {16},
	volume = {37},
	doi = {10.1002/adma.202419283},
	publisher = {Wiley},
	url = {http://dx.doi.org/10.1002/adma.202419283}
}

@article{Ohno2000,
	author = {Ohno, H. and Chiba, D. and Matsukura, F. and Omiya, T. and Abe, E. and Dietl, T. and Ohno, Y. and Ohtani, K.},
	journal = {Nature},
	title = {Electric-field control of ferromagnetism},
	year = {2000},
	issn = {1476-4687},
	month = Dec,
	number = {6815},
	pages = {944--946},
	volume = {408},
	doi = {10.1038/35050040},
	publisher = {Springer Science and Business Media LLC}
}

@article{Park2022,
	author = {Park, Nahee and Lee, Hyunkyung and Park, Jeongmin and Chau, Tuan Khanh and Kang, Hojin and Kang, Haeyong and Suh, Dongseok},
	journal = {NPG Asia Materials},
	title = {Charge carrier modulation in graphene on ferroelectric single-crystal substrates},
	year = {2022},
	issn = {1884-4057},
	month = July,
	number = {1},
	volume = {14},
	doi = {10.1038/s41427-022-00404-5},
	publisher = {Springer Science and Business Media LLC}
}

@article{Baltz2018,
	author = {Baltz, V. and Manchon, A. and Tsoi, M. and Moriyama, T. and Ono, T. and Tserkovnyak, Y.},
	journal = {Rev. Mod. Phys.},
	title = {Antiferromagnetic spintronics},
	year = {2018},
	month = {Feb},
	pages = {015005},
	volume = {90},
	doi = {10.1103/RevModPhys.90.015005},
	issue = {1},
	numpages = {57},
	publisher = {American Physical Society},
	url = {https://link.aps.org/doi/10.1103/RevModPhys.90.015005}
}

@article{Lee2021,
	author = {Lee, Kihong and Dismukes, Avalon H. and Telford, Evan J. and Wiscons, Ren A. and Wang, Jue and Xu, Xiaodong and Nuckolls, Colin and Dean, Cory R. and Roy, Xavier and Zhu, Xiaoyang},
	journal = {Nano Lett.},
	title = {Magnetic Order and Symmetry in the 2D Semiconductor CrSBr},
	year = {2021},
	issn = {1530-6984},
	month = apr,
	number = {8},
	pages = {3511--3517},
	volume = {21},
	comment = {csb shg T_Néel},
	doi = {10.1021/acs.nanolett.1c00219},
	publisher = {American Chemical Society},
	url = {https://doi.org/10.1021/acs.nanolett.1c00219}
}

@article{Jeong2026,
	author = {Jeong, Yeonsu and Wang, Honglei and Tordi, Pietro and Hu, Ying Chieh and Tamayo, Adrián and Wu, Bing and Ahn, Hyun Soo and Sofer, Zdenek and Jung, Jong Hoon and Bonini, Massimo and Samorì, Paolo},
	journal = {Advanced Science},
	title = {Boosting Ferroelectricity: 2D and Polymer Ferroelectric Hybrids Enabling Ambipolar Nonvolatile MoS2 Memory Transistor},
	number = {n/a},
	pages = {e76127},
	volume = {n/a},
	doi = {https://doi.org/10.1002/advs.76127},
	eprint = {https://advanced.onlinelibrary.wiley.com/doi/pdf/10.1002/advs.76127},
	url = {https://advanced.onlinelibrary.wiley.com/doi/abs/10.1002/advs.76127}
}

@article{Liu2025,
	author = {Liu, Ziqi and Sun, Yichang and Zhu, Chengfeng and Hong, Canyu and Gao, Yuchen and Sun, Zeyuan and Watanabe, Kenji and Taniguchi, Takashi and Wu, Shiwei and Chen, Zuxin and Gu, Pingfan and Ye, Yu},
	journal = {Phys. Rev. B},
	title = {Spin texture and tunneling magnetoresistance in atomically thin CrSBr},
	year = {2025},
	month = {Apr},
	pages = {L140417},
	volume = {111},
	comment = {csb},
	doi = {10.1103/PhysRevB.111.L140417},
	issue = {14},
	numpages = {8},
	publisher = {American Physical Society},
	url = {https://link.aps.org/doi/10.1103/PhysRevB.111.L140417}
}

@article{Telford2022,
	author = {Telford, Evan J. and Dismukes, Avalon H. and Dudley, Raymond L. and Wiscons, Ren A. and Lee, Kihong and Chica, Daniel G. and Ziebel, Michael E. and Han, Myung-Geun and Yu, Jessica and Shabani, Sara and Scheie, Allen and Watanabe, Kenji and Taniguchi, Takashi and Xiao, Di and Zhu, Yimei and Pasupathy, Abhay N. and Nuckolls, Colin and Zhu, Xiaoyang and Dean, Cory R. and Roy, Xavier},
	journal = {Nature Materials},
	title = {Coupling between magnetic order and charge transport in a two-dimensional magnetic semiconductor},
	year = {2022},
	issn = {1476-4660},
	month = jul,
	number = {7},
	pages = {754--760},
	volume = {21},
	comment = {csb},
	doi = {10.1038/s41563-022-01245-x},
	publisher = {Springer Science and Business Media LLC},
	refid = {Telford2022},
	url = {https://doi.org/10.1038/s41563-022-01245-x}
}

@article{Jia2025,
	author = {Jia, Zhiyan and Zhao, Mengfan and Chen, Qian and Tian, Yuxin and Liu, Lixuan and Zhang, Fang and Zhang, Delin and Ji, Yue and Camargo, Bruno and Ye, Kun and Sun, Rong and Wang, Zhongchang and Jiang, Yong},
	journal = {ACS Nano},
	title = {Spintronic Devices upon 2D Magnetic Materials and Heterojunctions},
	year = {2025},
	issn = {1936-086X},
	month = Mar,
	number = {10},
	pages = {9452--9483},
	volume = {19},
	doi = {10.1021/acsnano.4c14168},
	publisher = {American Chemical Society (ACS)}
}

@article{DalDin2024,
	author = {Dal Din, A. and Amin, O. J. and Wadley, P. and Edmonds, K. W.},
	journal = {npj Spintronics},
	title = {Antiferromagnetic spintronics and beyond},
	year = {2024},
	issn = {2948-2119},
	month = July,
	number = {1},
	volume = {2},
	doi = {10.1038/s44306-024-00029-0},
	publisher = {Springer Science and Business Media LLC}
}

@article{Neese2008,
	author = {Bret Neese and Baojin Chu and Sheng-Guo Lu and Yong Wang and E. Furman and Q. M. Zhang},
	journal = {Science},
	title = {Large Electrocaloric Effect in Ferroelectric Polymers Near Room Temperature},
	year = {2008},
	number = {5890},
	pages = {821-823},
	volume = {321},
	doi = {10.1126/science.1159655},
	eprint = {https://www.science.org/doi/pdf/10.1126/science.1159655},
	url = {https://www.science.org/doi/abs/10.1126/science.1159655}
}

@article{Naber2010,
	author = {Naber, Ronald C. G. and Asadi, Kamal and Blom, Paul W. M. and de Leeuw, Dago M. and de Boer, Bert},
	journal = {Advanced Materials},
	title = {Organic Nonvolatile Memory Devices Based on Ferroelectricity},
	year = {2010},
	number = {9},
	pages = {933-945},
	volume = {22},
	doi = {https://doi.org/10.1002/adma.200900759},
	eprint = {https://advanced.onlinelibrary.wiley.com/doi/pdf/10.1002/adma.200900759},
	url = {https://advanced.onlinelibrary.wiley.com/doi/abs/10.1002/adma.200900759}
}

@article{Gibertini2019,
	author = {Gibertini, M. and Koperski, M. and Morpurgo, A. F. and Novoselov, K. S.},
	title = {Magnetic 2D materials and heterostructures},
	journal = {Nature Nanotechnology},
	year = {2019},
	volume = {14},
	number = {5},
	pages = {408--419},
	month = may,
	issn = {1748-3395},
	refid = {Gibertini2019},
	url = {https://doi.org/10.1038/s41565-019-0438-6}
}

@article{Klein2018,
	author = {D. R. Klein and D. MacNeill and J. L. Lado and D. Soriano and E. Navarro-Moratalla and K. Watanabe and T. Taniguchi and S. Manni and P. Canfield and J. Fernández-Rossier and P. Jarillo-Herrero},
	journal = {Science},
	title = {Probing magnetism in 2D van der Waals crystalline insulators via electron tunneling},
	year = {2018},
	number = {6394},
	pages = {1218-1222},
	volume = {360},
	doi = {10.1126/science.aar3617},
	eprint = {https://www.science.org/doi/pdf/10.1126/science.aar3617},
	url = {https://www.science.org/doi/abs/10.1126/science.aar3617}
}

@article{Weisheit2007,
	author = {Martin Weisheit and Sebastian Fähler and Alain Marty and Yves Souche and Christiane Poinsignon and Dominique Givord},
	journal = {Science},
	title = {Electric Field-Induced Modification of Magnetism in Thin-Film Ferromagnets},
	year = {2007},
	number = {5810},
	pages = {349-351},
	volume = {315},
	doi = {10.1126/science.1136629},
	eprint = {https://www.science.org/doi/pdf/10.1126/science.1136629},
	url = {https://www.science.org/doi/abs/10.1126/science.1136629}
}

@article{LopezPaz2022,
	author = {López-Paz, Sara A. and Guguchia, Zurab and Pomjakushin, Vladimir Y. and Witteveen, Catherine and Cervellino, Antonio and Luetkens, Hubertus and Casati, Nicola and Morpurgo, Alberto F. and von Rohr, Fabian O.},
	journal = {Nature Communications},
	title = {Dynamic magnetic crossover at the origin of the hidden-order in van der Waals antiferromagnet CrSBr},
	year = {2022},
	issn = {2041-1723},
	month = aug,
	number = {1},
	volume = {13},
	comment = {csb},
	doi = {10.1038/s41467-022-32290-4},
	publisher = {Springer Science and Business Media LLC}
}

@article{Ningrum2020,
	author = {Ningrum, Vertikasari P. and Liu, Bowen and Wang, Wei and Yin, Yao and Cao, Yi and Zha, Chenyang and Xie, Hongguang and Jiang, Xiaohong and Sun, Yan and Qin, Sichen and Chen, Xiaolong and Qin, Tianshi and Zhu, Chao and Wang, Lin and Huang, Wei},
	journal = {Research},
	title = {Recent Advances in Two-Dimensional Magnets: Physics and Devices towards Spintronic Applications},
	year = {2020},
	issn = {2639-5274},
	month = Jan,
	volume = {2020},
	doi = {10.34133/2020/1768918},
	publisher = {American Association for the Advancement of Science (AAAS)}
}
 
\end{document}


\title{Supporting Material: \papertitle}
\newcommand{\fiztsz}{Department of Physics, Institute of Physics, Budapest University of Technology and Economics, Muegyetem rkp. 3., 1111 Budapest, Hungary}

\newcommand{\lenduletaffil}{MTA-BME Correlated van der Waals Structures Momentum Research Group, M\H{u}egyetem rkp. 3., H-1111 Budapest, Hungary}

\newcommand{\scaffil}{MTA-BME Superconducting Nanoelectronics Momentum Research Group, M\H{u}egyetem rkp. 3., H-1111 Budapest, Hungary}

\newcommand{\skaffil}{Institute of Quantum Science, Department of Physics, Inha University, 22212 Incheon, South Korea}

\newcommand{\cicaffil}{CIC nanoGUNE BRTA, 20018 Donostia--San Sebastian, Basque Country, Spain}

\newcommand{\ehuaffil}{Departamento de Polímeros y Materiales Avanzados: Física, Química y Tecnología, Facultad de Químicas (EHU), Apartado 1072, 20080, San Sebastián, Spain}

\newcommand{\cfmaffil}{Centro de Física de Materiales--Materials Physics Center (CFM-MPC), CSIC--UPV/EHU, Donostia--San Sebastián, 20018, Spain}

\newcommand{\ikeraffil}{Ikerbasque, Basque Foundation for Science, 48009 Bilbao, Spain}

\newcommand{\rugaffil}{Zernike Institute for Advanced Materials, University of Groningen, Groningen, Netherlands}

\newcommand{\isisaffil}{University of Strasbourg \& CNRS, ISIS \& icFRC, 8 allee Gaspard Monge, 67000 Strasbourg, France}

\newcommand{\inlaffil}{INL - International Iberian Nanotechnology Laboratory, 4715-330 Braga, Portugal}

\author{B\'alint \textsc{F\"ul\"op}}
\email{fulop.balint@ttk.bme.hu}
\affiliation{\fiztsz}

\author{Yeonsu \textsc{Jeong}}
\affiliation{\skaffil}

\author{Sofia \textsc{Ferreira-Teixeira}}
\affiliation{\cicaffil}
\affiliation{\inlaffil}

\author{Covadonga \textsc{\'Alvarez-Garc\'ia}}
\affiliation{\cicaffil}
\affiliation{\ehuaffil}

\author{Tam\'as \textsc{Prok}}
\affiliation{\fiztsz}
\affiliation{\lenduletaffil}
\affiliation{\rugaffil}

\author{Xiaomin \textsc{Guo}}
\affiliation{\cicaffil}

\author{Endre \textsc{T\'ov\'ari}}
\affiliation{\fiztsz}
\affiliation{\lenduletaffil}

\author{Jong Hoon \textsc{Jung}}
\affiliation{\skaffil}

\author{Luis \textsc{Hueso}}
\affiliation{\cicaffil}
\affiliation{\ikeraffil}

\author{Paolo \textsc{Samori}}
\affiliation{\isisaffil}

\author{Marco \textsc{Gobbi}}
\email{marco.gobbi@ehu.eus}
\affiliation{\cfmaffil}
\affiliation{\ikeraffil}

\author{P\'eter \textsc{Makk}}
\email{makk.peter@ttk.bme.hu}
\affiliation{\fiztsz}
\affiliation{\lenduletaffil}

\author{Szabolcs \textsc{Csonka}}
\affiliation{\fiztsz}
\affiliation{\scaffil} \date{\today}

\maketitle

\beginsupplement

\section{Fabrication details}
Pre-patterned contact electrodes were designed on Si/SiO$_2$ substrates using UV lithography. Ti/Au was deposited by evaporation with a thickness of 3 and 12\,nm, respectively, to minimise the thickness of the leads for easier device fabrication while maintaining a good electrical contact.

CrSBr flakes were prepared by mechanical exfoliation from a bulk CrSBr crystal purchased from HQ Graphene using a blue tape (Nitto SPV224) and transferred to a PDMS film (Gel-pak).
The thickness of each flake was estimated based on optical contrast.
The corerlation between optical contrast and thickness was calibrated using optical micrographs of previous reference CrSBr flakes of thicknesses known from AFM measurements.
Select CrSBr flakes were transferred onto the pre-patterned Ti/Au electrodes using the standard dry transfer stamping technique.
Both the exfoliation and stamping were performed in ambient conditions.

p(VDF-TrFE) (Sigma-Aldrich, product code: 900904) was dissolved in 2-butanone to prepare a 6 wt.\% solution.
The solution was spin-coated onto the CrSBr devices and dried at 100$^\circ$C for 1 min under a nitrogen (N$_2$) atmosphere in a glove box.
The films were subsequently annealed at 142$^\circ$C for 2 h inside the N$_2$-filled glove box to induce crystallization of the ferroelectric $\beta$ phase.
A top-gate electrode was fabricated by photolithography (laser writer LW405B, Microtech; AZ1505 photoresist), thermal evaporation, and a lift-off.
The fabricated devices were finally annealed at 100$^\circ$C for 2 h under N$_2$ to improve contact quality.

The samples were diced to ca.\ $5\unit{mm} \, \times \, 5\unit{mm}$ and glued in a ceramic chip carrier using silver conducting paint.
Finally, $25\unit{\mu m}$ thick Au bonding wires were attached manually to the contact pads and fixed by using Micromax 6838 silver epoxy without curing.

\section{Polymer characterization}
The capacitance of the p(VDF-TrFE) films was measured by a precision LCR meter (Agilent 4284A) at 1\,kHz under ambient conditions in the dark.
The polarization vs.\ gate voltage hysteresis loops were measured by a ferroelectric tester (TF analyzer 1000) at 100\,Hz using a top-electrode with an area of $0.36\unit{mm^2}$.

\begin{figure}
\includegraphics[width=\textwidth]{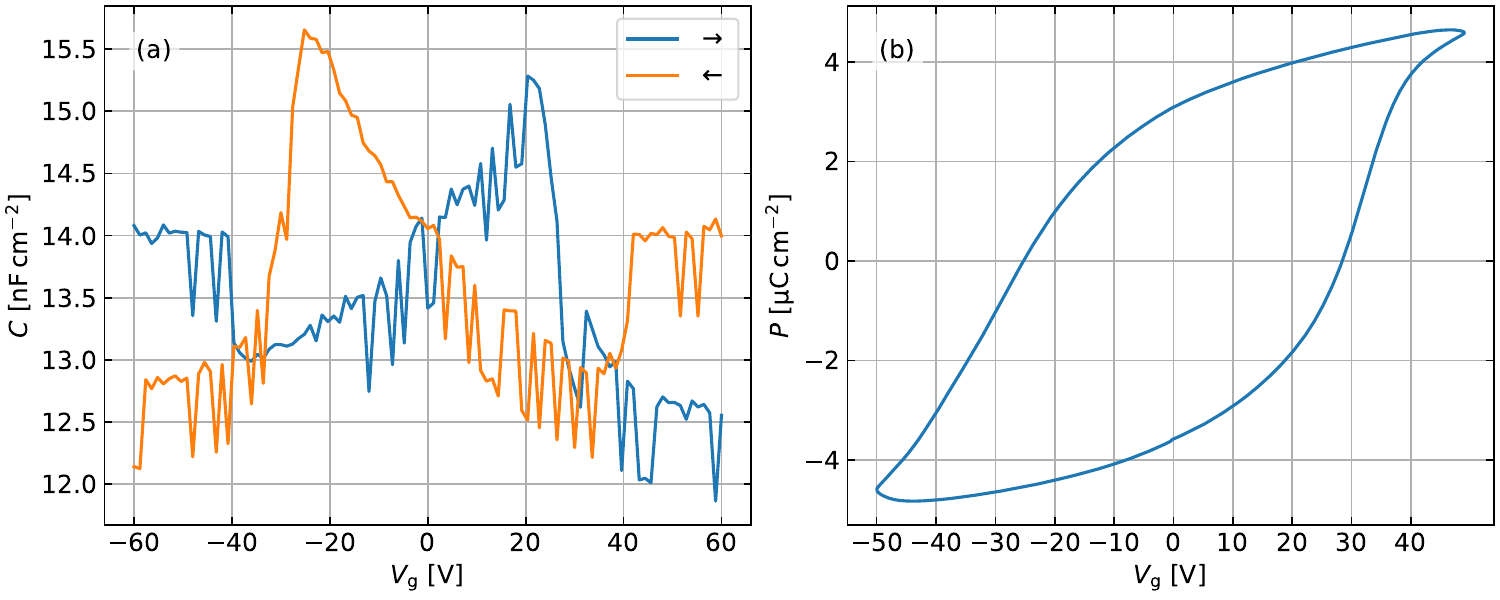}
\caption{
Polymer characterization
(a)
Capacitance measurement,
(b)
polarization--electric field measurement of a 500\,nm thick polymer film.
}
\label{fig:sm_polymer}
\end{figure}

The polymer characterization measurements are shown in Fig.\,\ref{fig:sm_polymer}.
In panel (a), the capacitance is shown vs.\ the gate voltage for the up (blue) and down (orange) sweep directions for a 500\,nm thick polymer layer, which is the same thickness as used in the CrSBr samples.
A small peak can be observed near the dipole flip transition.
The mean measured capacitance is $C = 13.5 \pm 0.75 \unit{nF\, cm^{-2}}$, which corresponds to $\varepsilon_\tn{r} = C \, d / \varepsilon_0 = 7.6 \pm 0.4$.

\section{Transconductance measurements}
Electrical measurements were performed in a liquid helium system with a variable temperature insert (VTI) at various temperatures between $1.5\unit{K}$ and $300\unit{K}$.
The preparation of the polymer dipole state was done at room temperature requiring warming up the sample, applying a gate voltage sweep with a simultaneous transconductance measurement, and cooling it down back to low temperature for the MR measurements.
During the voltage sweep, $\vtg$ was ramped from zero to $\vx$ at a rate of ca.\ $1\unit{V/s}$, settled for 1\,s, then ramped back to zero.
Although CrSBr is known to be air stable, application of a top gate voltage comparable to the coercive voltage of the polymer in open air caused sample degradation and loss of conductance of the CrSBr.
Therefore, all the transconductance measurements presented here were performed in He environment in the VTI.
A Basel Instruments SP927 DAC instrument was used to apply the source--drain voltage $\vsd$ and the top gate voltage $\vtg$.
The current was recorded by a Basel Instruments SP983 I/V converter and a Keithley 2001 multimeter in DC voltage mode.

 Extraction of the coercive field with error is based on subsequent transconductance measurements.
Subsequent curves scatter much more in the positive region than in the negative.
This scattering was used to estimate the uncertainty of the values of the coercive voltage on each side.
As indicated b steps 2, 3, 5, and 6 in the paper, the total onset of the \up{} (\down) state is only achieved at voltages where the conductance curve reaches the high (low) saturation value.
Therefore, to determine the coercive voltages $\vcp$ and $\vcn$, we take the voltages of the curves in the proximity of the saturation values: at $0.68\,G_\tn{mean}$ and $1.21\,G_\tn{mean}$ for the negative and positive coercive voltages, respectively.
In the negative region, all the curves fall within $\vcn = 32.9 \pm 1.7 \unit{V}$ except the very first curve at considerably higher voltage, which can be attributed to some initial effect.
In the positive region, $\vcp = 31.8 \pm 3.5 \unit{V}$ can be read.
 
\section{Further magneto-resistance measurements}
\begin{figure*}
\includegraphics[width=\textwidth]{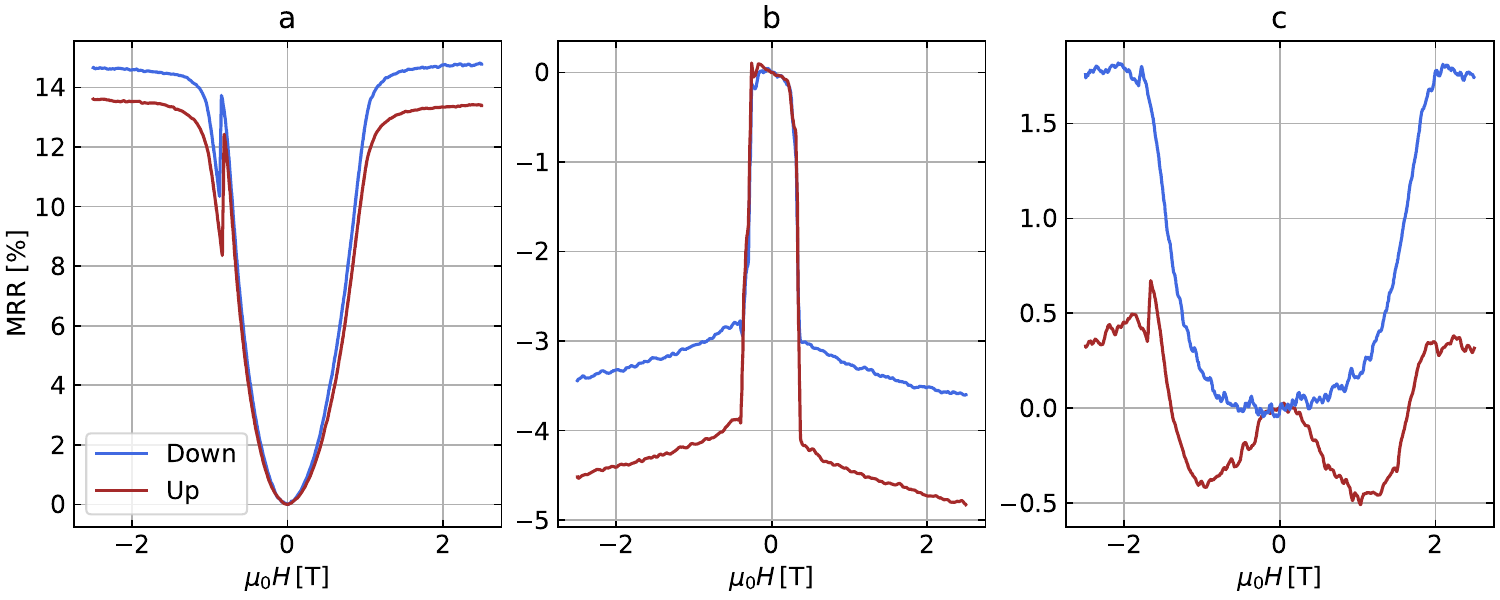}
\caption{
Comparison of MRR curves of the main sample along different magnetic axes at 15\,K:
a: in-plane intermediate axis,
b: in-plane easy axis,
c: out-of-plane hard axis.
The red (blue) lines correspond to the \up{} (\down{}) polymer dipole states.
}
\label{smfig:mrr_A_full}
\end{figure*}

In Fig.\,\ref{smfig:mrr_A_full} MRR data on the main sample with magnetic field along different crystallographic axes is shown.
The measurements were done at 15\,K.
In panel (a), MR in magnetic field along the in-plane intermediate axis ($a$) is shown, which exhibit no important difference between the \up{} (red) and \down{} (blue) dipole states.
With increasing magnetic field, the spins start to cant toward the magnetic field direction leading to a smoothly changing MRR signal, which saturates at the saturation field $\bsat$, where the system enters a fully polarized state.
In panel (b), MR in magnetic field along the easy axis ($b$) is shown.
The line shape shows a plateau in low magnetic fields with a sharp jump at the saturation field corresponding to the spin flip transition \cite{Telford2022}.
Here, the magnitude of the MRR of of the \up{} state is ca.\ four times larger than that of the \down{} state.
In panel (c), MR in out-of-plane magnetic field is shown, which is also discussed in the main text.
The saturation field $\bsat$ is different along the crystallographic axes, but they are measured the same for the \up{} and \down{} dipole states.

\begin{figure}
\includegraphics[height=2cm]{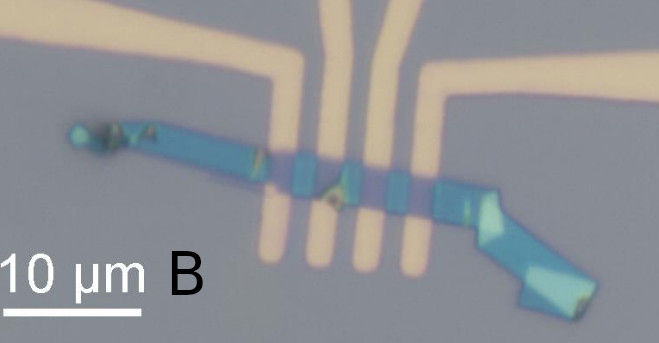}
\hspace{1 cm}
\includegraphics[height=2cm]{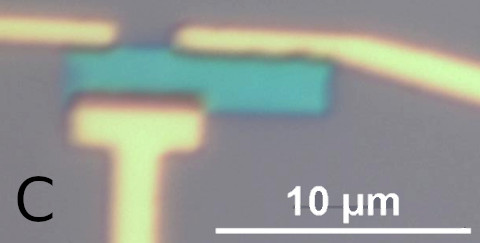}
\caption{
Optical micrographs of the second sample prior to polymer coating and the bare reference sample in its final form, both contacted by individual metallic contacts.
Gray: bare surface of the SiO$_2$ substrate,
green: exfoliated CrSBr flake with some of its parts folded back (light green),
yellow: pre-patterned Ti/Au electrodes.
The scale bar is 10\,\textmu{}m.
Flake thicknesses were estimated using AFM measurements.
Second sample (b): 4--6 layers.
Reference sample (c): 8 layers, same as the main sample.
}
\label{fig:micropt_e8}
\end{figure}

\begin{figure*}
\includegraphics[width=\textwidth]{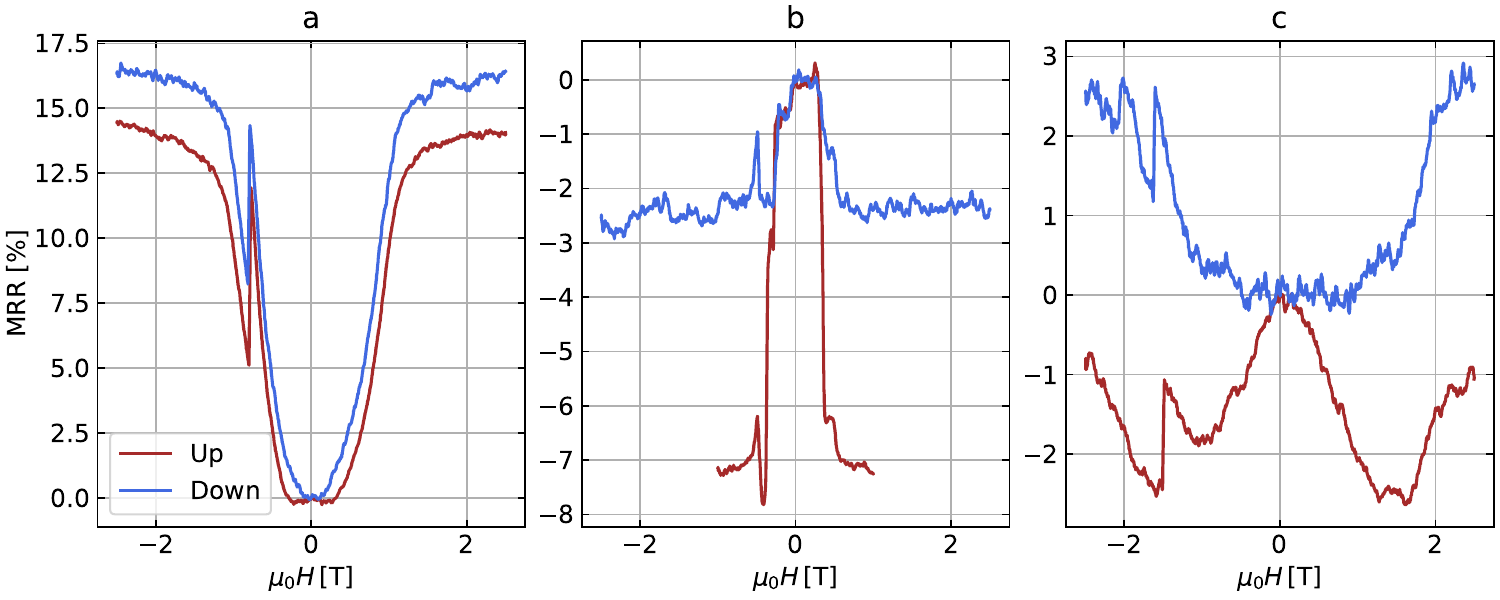}
\caption{
Comparison of MRR curves of the second sample along different magnetic axes at 15\,K:
a: in-plane intermediate axis,
b: in-plane easy axis,
c: out-of-plane hard axis.
The red (blue) lines correspond to the \up{} (\down{}) polymer dipole states.
The observed effects in each axis are similar as for the main sample.
}
\label{smfig:mrr_B_full}
\end{figure*}

A second sample is created following the same design, an optical micrograph of which is shown in Fig.\,\ref{fig:micropt_e8}.
In Fig.\,\ref{smfig:mrr_B_full}, MRR data on the second sample is presented with magnetic field along different crystallographic axes, also measured at 15\,K.
The observed features are similar to that of the main sample but noisier.

\begin{figure}
\includegraphics[width=\textwidth]{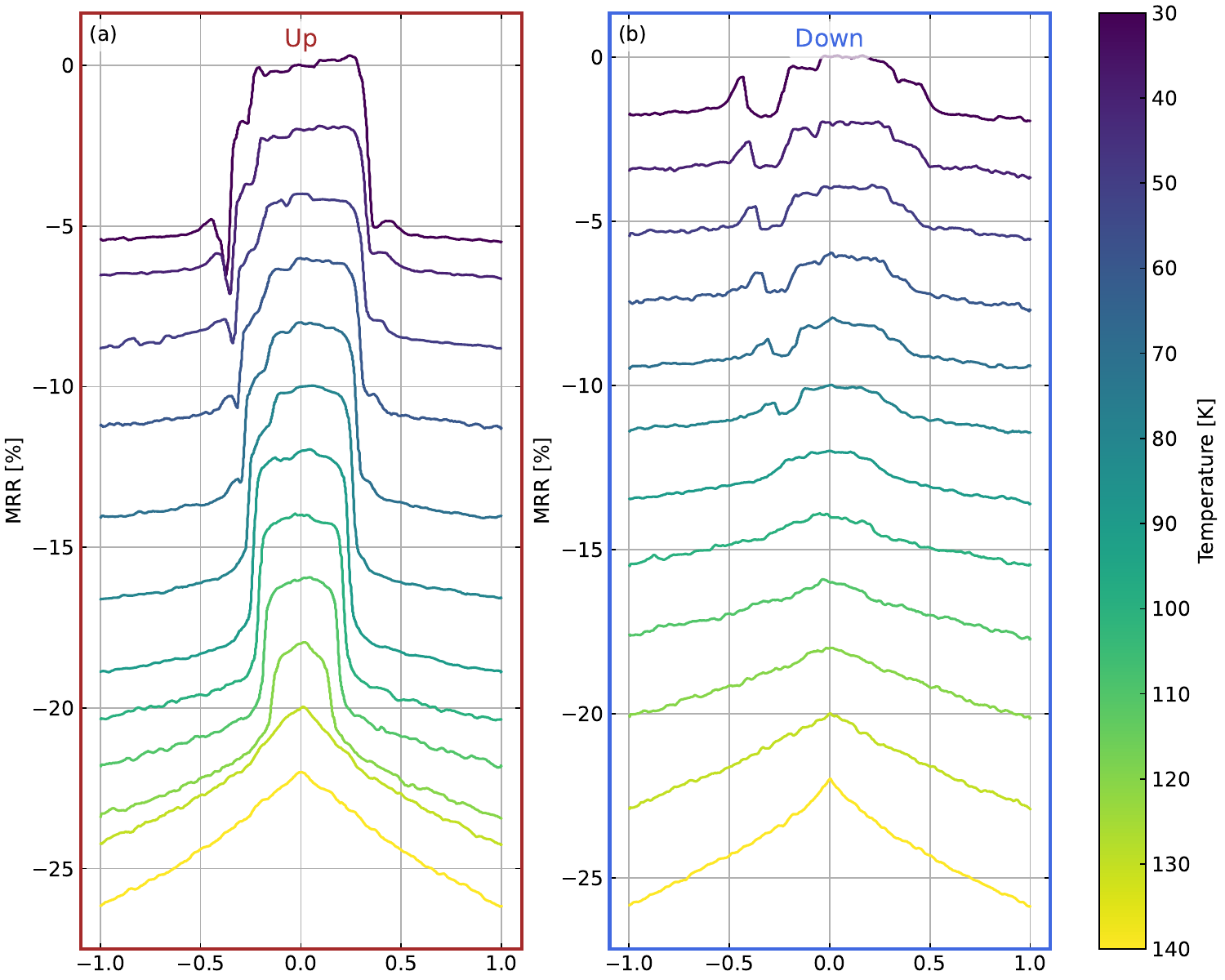}
\caption{
MRR of the second sample with the magnetic field along the easy axis ($b$) for the \up{} (a) and \down{} (b) dipole states at various temperatures indicated in the color scale.
The difference in the MRR signal presented in \ref{smfig:mrr_B_full} persists up to $\tneel$.
}
\label{fig:e8MRR}
\end{figure}

 A summary of MRR plots is shown in Fig.\,\ref{fig:e8MRR} on the second sample with magnetic field along the easy axis ($a$) at various temperatures.
On panel (a), the \up{} state is presented, which shows a consistently larger MRR at $\bsat$ than in the case of the \down{} state presented in panel (b).
Also, the latter exhibit more pronounced inner jumps in the plateau region, which correspond to multidomain steps.
Due to the overall lower MRR in the \down{} state, the features stemming from the AFM phase are more recognizable up to $\tneel$.

\bibliography{cited}